# MTBF-33: A multi-temporal building footprint dataset for 33 counties in the United States (1900 – 2015)

**Johannes H. Uhl**[1,2,3*], **Stefan Leyk**[2,3]

[1]University of Colorado Boulder, Cooperative Institute for Research in Environmental Sciences (CIRES) 216 UCB, Boulder, CO-80309, USA.
[2]University of Colorado Boulder, Institute of Behavioral Science, 483 UCB, Boulder, CO-80309, USA.
[3]University of Colorado Boulder, Department of Geography, 260 UCB, Boulder, CO-80309, USA.
*Corresponding author: Johannes.Uhl@colorado.edu

**Abstract:** Despite abundant data on the spatial distribution of contemporary human settlements, historical data on the long-term evolution of human settlements at fine spatial and temporal granularity is scarce, limiting our quantitative understanding of long-term changes of built-up areas. This is because commonly used mapping methods (e.g., image classification) and suitable data sources (i.e., aerial imagery, multi-spectral remote sensing data, LiDAR) have only been available in recent decades. However, there are alternative data sources such as cadastral records that are digitally available, containing relevant information such as building age information, allowing for an approximate, digital reconstruction of past building distributions. We conducted a non-exhaustive search of open and publicly available data resources from administrative institutions in the United States and gathered, integrated, and harmonized cadastral parcel data, tax assessment data, and building footprint data for 33 counties, wherever building footprint geometries and building construction year information was available. The result of this effort is a unique dataset which we call the Multi-Temporal Building Footprint Dataset for 33 U.S. Counties (MTBF-33). MTBF-33 contains over 6.2 million building footprints including their construction year, and can be used to derive retrospective depictions of built-up areas from 1900 to 2015, at fine spatial and temporal grain and can be used for data validation purposes, or to train statistical learning approaches aiming to extract historical information on human settlements from remote sensing data, historical maps, or similar data sources. MTBF-33 is available at http://doi.org/10.17632/w33vbvjtdy.

**Value of the data**

- Open and publicly accessible data on building age are scarce. Our data scraping, integration, and harmonization effort aims to fill this gap in the data landscape.
- Knowledge of the construction year of individual buildings allows for creating spatially and temporally fine-grained depictions of past built-up surfaces.
- Such spatial-historical data may serve as calibration and validation data for urban change models and for historical (and more recent) human settlement datasets (cf. See et al. 2022), as well as for historical population downscaling efforts.
- Moreover, such data are very useful to be employed as auxiliary data for the automated training data generation for data-intensive (e.g. deep learning) computer vision models to automatically extract urban change signals from remote sensing data or historical maps.

- Lastly, these data enable historical analyses of the building stock in 33 U.S. counties, encompassing the whole state of Massachusetts, as well as several urban areas of different settlement age and characteristics, such as Boston, Charlotte, and Minneapolis.
- These data are highly valuable for urban planners, remote sensing analysts, historians, demographers, and data scientists working in the context of urban land use change and (sub)urbanization, as they provide rare insight into the long-term dynamics of built-up areas at very high spatial and temporal detail.

**Data description**

We collected open and publicly available data resources from the web from administrative, county- or state-level institutions in the United States and integrated and harmonized cadastral parcel data, tax assessment data, and building footprint data for 33 counties, where building footprint data and building construction year information ("year built") was available. The result of this effort is a unique dataset called the Multi-Temporal Building Footprint Dataset for 33 U.S. Counties (MTBF-33, Uhl & Leyk 2022, available at http://dx.doi.org/10.17632/w33vbvjtdy). MTBF-33 contains over 6.2 million building footprints including their construction year, and is available as polygonal geospatial vector data in 33 ESRI Shapefiles, projected into Albers equal area conic projection for the contiguous USA (USGS version, SR-ORG:7480[1]), organized per county. Figure 1 shows small subsets of the MTBF-33 dataset for selected regions.

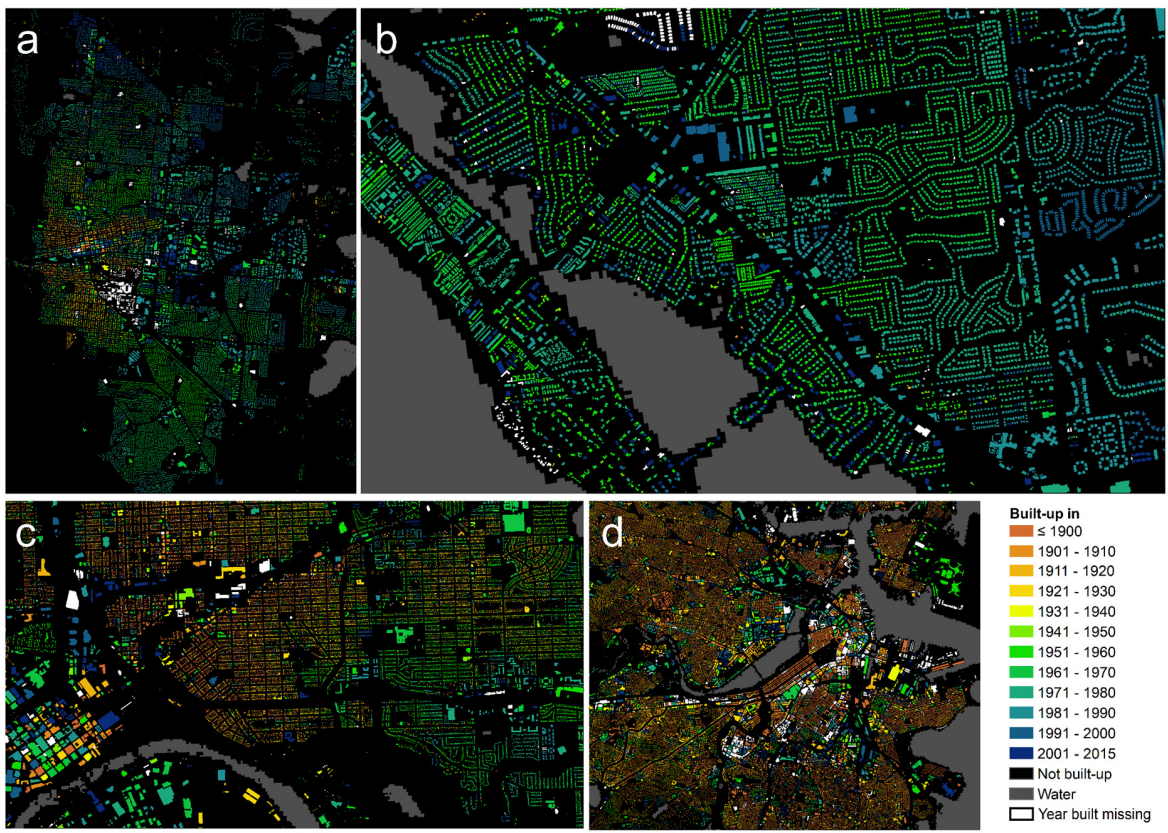

**Figure 1.** MTBF-33 multi-temporal building footprint data examples, shown for (a) Boulder (Colorado) (b) Sarasota (Florida), (c) Boston (Massachusetts), and (d) Minneapolis (Minnesota).

---

[1] https://spatialreference.org/ref/sr-org/7480/

Moreover, Figure 2 shows a small subset of the data for most of the 33 U.S. counties, illustrating the variability in the data and their coverage across different geographic settings.

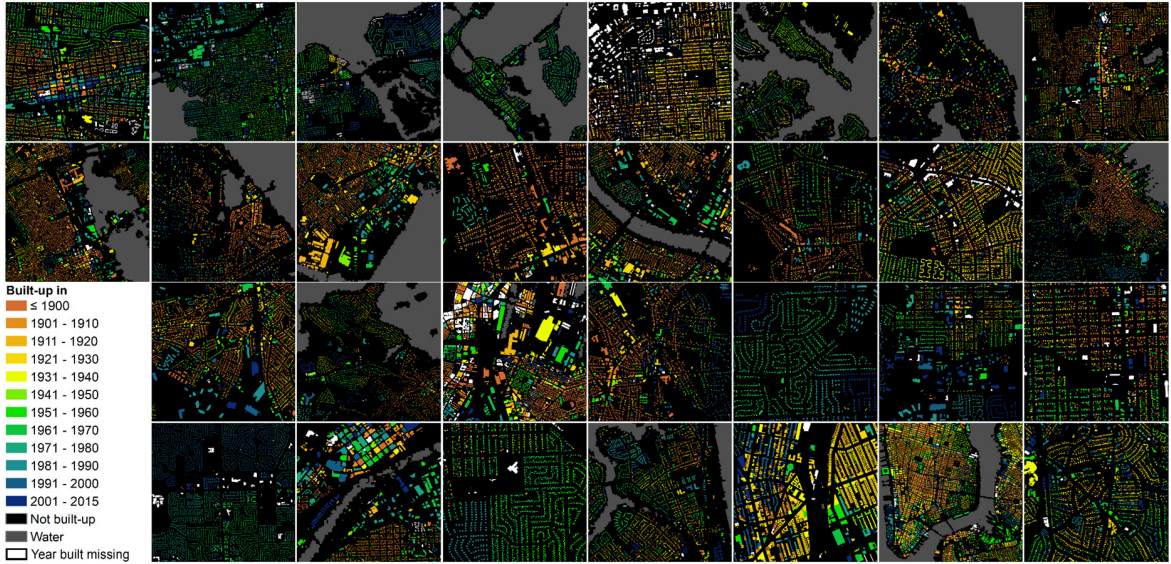

**Fig. 2. MTBF-33 multi-temporal building footprint dataset, showing examples for most of the 33 counties covered in MTBF-33. Water mask in grey derived from GHS-BUILT R2018A (epoch 2014, Florczyk et al. 2019). Counties are sorted by their FIPS in the same order as shown in Table 1 (upper left: Boulder County, lower right: Mecklenburg county). Not shown are Queens and Richmond counties (New York City).**

As can be seen in Figures 1 and 2, there are several buildings without year built attribute (white color). We report the year built attribute completeness for each of the 33 counties in Table 1. Moreover, Table 1 shows some basic year built statistics, illustrating the variety in temporal coverage of the data.

**Table 1. Overview and statistics of the 33 counties covered in MTBF-33.**

| County FIPS | County Name | State | Buildings w/ valid year built | Total buildings | Percent complete | Year built minimum | Year built maximum | Year built mean | Year built median |
|---|---|---|---|---|---|---|---|---|---|
| 08013 | Boulder | Colorado | 76929 | 80255 | 95.9 | 1858 | 2014 | 1968 | 1971 |
| 12057 | Hillsborough | Florida | 410076 | 421046 | 97.4 | 1842 | 2014 | 1981 | 1984 |
| 12081 | Manatee | Florida | 154416 | 173173 | 89.2 | 1870 | 2013 | 1980 | 1982 |
| 12115 | Sarasota | Florida | 194043 | 198685 | 97.7 | 1877 | 2013 | 1981 | 1981 |
| 18163 | Vanderburgh | Indiana | 93797 | 108798 | 86.2 | 1810 | 2015 | 1951 | 1950 |
| 24005 | Baltimore | Maryland | 281216 | 308933 | 91.0 | 1676 | 2015 | 1958 | 1958 |
| 25001 | Barnstable | Massachusetts | 172542 | 185818 | 92.9 | 1626 | 2014 | 1961 | 1971 |
| 25003 | Berkshire | Massachusetts | 82036 | 89790 | 91.4 | 1650 | 2015 | 1937 | 1950 |
| 25005 | Bristol | Massachusetts | 225156 | 233704 | 96.3 | 1500 | 2013 | 1948 | 1957 |
| 25007 | Dukes | Massachusetts | 18758 | 23524 | 79.7 | 1660 | 2014 | 1963 | 1980 |
| 25009 | Essex | Massachusetts | 224351 | 270398 | 83.0 | 1600 | 2014 | 1937 | 1947 |
| 25011 | Franklin | Massachusetts | 43436 | 50209 | 86.5 | 1666 | 2013 | 1936 | 1951 |
| 25013 | Hampden | Massachusetts | 192281 | 207195 | 92.8 | 1600 | 2015 | 1947 | 1953 |
| 25015 | Hampshire | Massachusetts | 69505 | 77982 | 89.1 | 1629 | 2014 | 1947 | 1960 |
| 25017 | Middlesex | Massachusetts | 460722 | 500047 | 92.1 | 1600 | 2015 | 1942 | 1950 |
| 25019 | Nantucket | Massachusetts | 13547 | 13971 | 97.0 | 1640 | 2011 | 1962 | 1983 |
| 25021 | Norfolk | Massachusetts | 216150 | 242631 | 89.1 | 1500 | 2015 | 1944 | 1951 |
| 25023 | Plymouth | Massachusetts | 207264 | 230788 | 89.8 | 1600 | 2015 | 1950 | 1962 |
| 25025 | Suffolk | Massachusetts | 106037 | 109876 | 96.5 | 1637 | 2015 | 1924 | 1920 |
| 25027 | Worcester | Massachusetts | 317302 | 344307 | 92.2 | 1650 | 2015 | 1948 | 1957 |
| 27003 | Anoka | Minnesota | 128498 | 135307 | 95.0 | 1852 | 2015 | 1977 | 1979 |
| 27019 | Carver | Minnesota | 40488 | 41768 | 96.9 | 1816 | 2015 | 1969 | 1984 |
| 27037 | Dakota | Minnesota | 145903 | 163179 | 89.4 | 1832 | 2014 | 1973 | 1978 |
| 27053 | Hennepin | Minnesota | 380301 | 387856 | 98.1 | 1843 | 2010 | 1955 | 1956 |
| 27123 | Ramsey | Minnesota | 239544 | 245279 | 97.7 | 1850 | 2015 | 1946 | 1951 |
| 27163 | Washington | Minnesota | 86216 | 95014 | 90.7 | 1742 | 2015 | 1973 | 1983 |
| 34025 | Monmouth | New Jersey | 206624 | 212951 | 97.0 | 1684 | 2015 | 1961 | 1963 |
| 36005 | Bronx | New York | 102658 | 103865 | 98.8 | 1780 | 2015 | 1941 | 1931 |
| 36047 | Kings | New York | 329283 | 331813 | 99.2 | 1800 | 2015 | 1931 | 1925 |
| 36061 | New York | New York | 45322 | 46209 | 98.1 | 1765 | 2014 | 1921 | 1910 |
| 36081 | Queens | New York | 454506 | 457628 | 99.3 | 1661 | 2015 | 1939 | 1935 |
| 36085 | Richmond | New York | 138609 | 140050 | 99.0 | 1665 | 2014 | 1962 | 1969 |
| 37119 | Mecklenburg | North Carolina | 402242 | 418056 | 96.2 | 1792 | 2015 | 1980 | 1984 |

**Experimental design, materials and methods**

    **1) Data creation**

We manually collected cadastral parcel data, tax assessment data, and building footprint data from publicly and openly available web resources, such as from state-level or county-level administrative GIS or spatial data resources. We used open data portals such as https://hub.arcgis.com to identify counties or states where (a) both parcel and building footprint data is available, and (b) parcel data or joinable tax assessment data contains information on the year when structures have been established (year built). We identified 33 counties that satisfied these criteria and where the completeness of the building footprint data and the year built attribute was acceptable (see Table 1).

In counties where the year built information was contained in separate tax assessment datasets, we first joined the tax assessment data to the parcel data. Then, we integrated the parcel data and building footprint data. This integration was done through a spatial join operation, in order to transfer the year built attribute from the parcel polygon features to the building footprint features contained within the cadastral parcel boundaries. This spatial assignment was based on a majority-area criterion in order to account for certain levels of spatial offsets between parcel and building footprint data. Such offsets may exist due to different data acquisition methods: While parcel boundaries are typically measures using terrestrial or Global Navigation Satellite System (GNSS)-based land surveying technologies, building footprint data may be obtained through automatic segmentation of LiDAR data or by digitization in aerial imagery.

As a result of these spatial joins, the year built attribute was transferred to the building footprint features. For these processes, we used the GeoPandas[2] and ESRI ArcPy[3] python package. We then harmonized and cleaned the data. This cleaning process involved the identification of non-plausible year built values (e.g., < 1500). Missing or non-plausible year built values were set to 0. Importantly, any property-, building-, or individual-level data other than the year built attribute was removed, so that the MTBF-33 data exclusively consists of building footprint geometries and their construction year. The resulting polygonal, geospatial vector data represent building footprints for 33 counties in the conterminous United States, allowing for the reconstruction of spatially and temporally fine-grained depictions of built-up surfaces (i.e., building level, annual resolution).

While contemporary building footprint data is available at high levels of accuracy (Microsoft 2018), data on the historical distributions of the U.S. building stock is very scarce, in particular for time periods earlier than the 1970s or 1980s, when remote-sensing-based, digital earth observation data became accessible. Thus, despite representing only about 1% of the U.S. counties, this unique dataset covers more than 40,000 km² and more than 6,000,000 cadastral parcels, and fills an important gap in the geospatial data landscape. The MTBF-33 dataset was collected in 2016-2017 and since then, MTBF-33 has been employed by the authors for different purposes, including the validation of global remote-sensing-based multi-temporal built-up surface data (Uhl et al., 2016, Leyk et al. 2018, Uhl et al. 2018), the validation of historical settlement data derived from property databases (Leyk & Uhl 2018, Uhl et al. 2021), to automatically generate training data for urban change detection based on Landsat time series data (Uhl & Leyk 2020), to assess the sensitivity of Landsat time series to urban changes (Uhl & Leyk 2017), and for training data generation used by computer vision models to extract settlement patterns from historical topographic maps (Uhl et al. 2017, Uhl et al. 2018).

---

[2] https://geopandas.org
[3] https://www.esri.com/pythonlibraries

**2) Validation and uncertainty assessment**

The validation of data that entail advancements in quality and accuracy compared to existing data products is always challenging. We evaluated MTBF-33 through two approaches. First, we compared the dataset for all 33 counties with the Microsoft Building Footprint (MSBF) dataset (Microsoft 2018) for the building footprints existing in 2015. Second, we carried out a visual comparison between the MTBF-33 and urban extents as found in historical topographic maps.

**2.1. Agreement assessment between MTBF-33 and Microsoft building footprint data**

Microsoft's building footprint dataset (released in 2018) has a US-wide coverage and has been extracted from Microsoft Bing imagery using a deep-learning-based computer vision method. While the acquisition years of the Bing imagery are likely to vary across the United States, we assume that MSBF represents the U.S. building stock approximately in 2015. In order to evaluate the MTBF-33 quantitatively, we created gridded binary layers (i.e., built-up versus not built-up) for 2015 for both MTBF-33 and MSBF, in a spatial grid of 250m x 250m, based on an area majority rule, for each of the 33 counties covered in MTBF-33. Based on these gridded surfaces, we established confusion matrices per county, used to calculate various agreement measures to assess agreement between the two binary layers (Table 2). While some of these measures have been criticized due to some limitations, e.g., if class proportions are imbalanced (Pontius & Millones 2011, Stehman & Wickham 2020), a cross-section through all those measures represents a reliable assessment basis. We carried out the agreement assessment across all 33 counties, as well as separately for higher-density and lower-density counties. This stratification was done based on the MSBF built-up surface density per county, using the median county-level built-up surface density as a threshold.

As can be seen in Table 2, when using all 33 counties for map comparison, all accuracy measures show high agreement between the two layers (between 86.5% based on Kappa and 95.6% based on Precision). Higher accuracy is observed for high-density counties compared to the low-density counties. The notable difference in Recall (0.99 and 0.85, respectively) indicates that omission errors are higher in low-density counties, possibly because MSBF identifies structures that are not part of the county assessor's building stock database (e.g., barns), especially in more rural settings. Thus, MTBF-33 has no built structures at those locations, resulting in higher proportions of false negatives. This effect propagates into the other measures, resulting in reduced F-measure and Kappa index for lower-density counties. However, even in lower-density counties no accuracy measure is less than 83% indicating high levels of agreement between the two datasets. Moreover, there may be slight temporal gaps between MTBF-33 and MSBF due to the image acquisition years of the imagery underlying the MSBF data, due to the heterogeneous levels of temporal coverage of the MTBF-33 data (see built year statistics in Table 1), and due to the vagueness in the definition of the construction year in MTBF-33.

**Table 2. Results of the agreement assessment between MTBF-33 (in 2015) and Microsoft building footprint data.**

| Agreement measure | All counties | Higher-density counties | Lower-density counties |
|---|---|---|---|
| Overall accuracy | 0.933 | 0.969 | 0.919 |
| Precision | 0.956 | 0.960 | 0.954 |
| Recall | 0.900 | 0.990 | 0.853 |
| F-measure | 0.928 | 0.974 | 0.901 |
| Kappa index | 0.865 | 0.935 | 0.833 |

**2.2. Qualitative evaluation of MTBF-33 data against historical maps**

We carried out a visual comparison between the MTBF-33 and urbanized extents as shown in historical topographic maps of the U.S. Geological Survey (USGS) historical topographic map collection (HTMC[4]) (Allord et al. 2014). To do so, we created historical binary layers of the MTBF-33 to match the publication dates of the historical map sheets. Figure 3 shows historical map sheets for Boulder, Colorado, and the respective extracted MTBF-33 binary layers for 1904, 1957 and 1984. As can be seen, the spatial representations of built-up / urbanized areas generally agree. Agreement is highest within urban areas as shown by denser building blocks along roads in 1904, in red in 1957 and grey in 1984, even though MTBF-33 shows much finer spatial detail of the built-up areas. Outside the urbanized areas, the 1904 and 1957 maps show detailed building symbols along roads, many of which are also visible in the MTBF-33 layer. Some discrepancies can be seen in the North-west part of the 1984 map, where some buildings are visible in MTBF-33 but not in the historical map. This is due to the level of cartographic generalization used in the 1984 map sheet (scale 1:100,000, whereas the 1904 and 1957 maps are at scale 1:62,500) which may not include individual building footprints outside urban extents.

Moreover, such discrepancies may be due to temporal uncertainty in the historical maps (i.e., temporal gap between land surveying or field check, and map edition / publishing year) and due to the vagueness of the construction year information in MTBF-33. It is unknown whether the date on record reflects the beginning or the end of the building construction phase, and how long the construction phase endured. Moreover, the construction year could be an estimate, and buildings may be missing in MTBF-33 because of incomplete records or missing built year information. However, the visual similarity for the two historical map sheets combined with the quantitative agreement assessment against Microsoft's building footprint dataset provide strong confidence that the built-up surfaces in MTBF-33 are very plausible and accurate with some local variations in completeness.

---

[4] Available at https://ngmdb.usgs.gov/topoview/

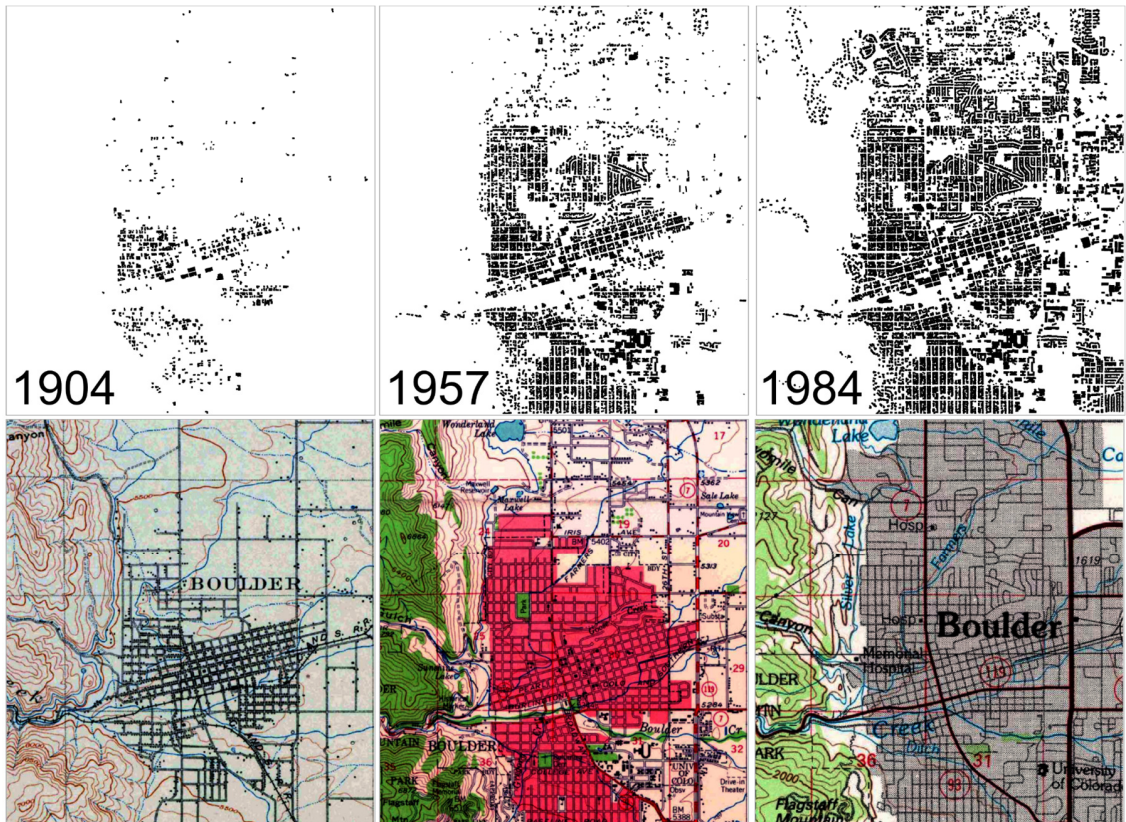

**Figure 3. Comparison of retrospective building footprint distributions to historical maps for Boulder, Colorado, in 1904 (scale 1:62,500), 1957 (scale 1:62,500), and 1984 (scale 1:100,000). Map source: USGS-HTMC.**

There are uncertainties in MTBF-33 that are very difficult to measure. For example, buildings that have been torn down or were destroyed are not included in the dataset. The respective parcels may have become vacant land or a new structure may have been built. As a consequence, there is survivorship bias in construction year information which increases as as we go back in time. There are few studies that report on or measure survivorship bias in settlement layers as this requires access to historical versions of cadastral data or demolition records. For example, McShane et al. (2022) used historical demolition data for Colorado and found that survivorship bias had limited impact on the resulting settlement layers of less than 2%. Note that while we retain all plausible year built values from the scraped source data, we constrain the temporal coverage to the period 1900 – 2015. We discourage data users to use MTBF-33 to create snapshots of built-up surfaces < 1900, as the survivorship bias may be very large.

**Specifications table**

| Subject | Geography |
|---|---|
| **Specific subject area** | Urban change detection, long-term land development, built environment, human settlements |
| **Type of data** | Geospatial vector data |
| **How the data were acquired** | Data were collected, integrated, and harmonized from web-based, open and publicly available sources published by local or regional governmental organizations, such as county or state governments. |
| **Data format** | **Raw:** ESRI Shapefile, ESRI File Geodatabase, Excel spreadsheets, CSV files.<br>**Filtered:** ESRI Shapefile |
| **Description of data collection** | We identified U.S. counties or states that provide building footprint data and cadastral parcel data attributed with building construction year information. In a non-exhaustive search we identified 33 U.S. counties where these criteria were met. We integrated and harmonized these data to create geospatial vector datasets holding over 6.2 million building footprints attributed with their construction year. |
| **Data source location** | Source data was collected in 2016 from the following resources:<br>ftp://ftp.co.ramsey.mn.us/GISdata/ (last accessed: 2016-03-01)<br>ftp://ftp.lmic.state.mn.us/pub/data/elevation/lidar/county/ (last accessed: 2016-03-01)<br>ftp://ftp1.fgdl.org/pub/state/ (last accessed: 2016-03-01)<br>ftp://gisdata.co.anoka.mn.us/ (last accessed: 2016-03-01)<br>http://bostonopendata.boston.opendata.arcgis.com/datasets/f3d274161b4a47aa9acf48d0d04cd5d4_3 (last accessed: 2016-03-01)<br>http://data.evansvilleapc.opendata.arcgis.com/datasets/0f1a8007227641d394f4170acba8aa67_1 (last accessed: 2016-03-01)<br>http://maps.co.mecklenburg.nc.us/openmapping/data.html (last accessed: 2022-03-02)<br>http://opendata.arcgis.com/datasets/611eb2cad81a4089afa188233e6b6dd1_0 (last accessed: 2022-03-02)<br>http://www.co.carver.mn.us/GIS (last accessed: 2022-03-02)<br>http://www.co.dakota.mn.us/homeproperty/propertymaps/pages/default.aspx (last accessed: 2022-03-02)<br>http://www.co.ramsey.mn.us/is/gisdata.htm (last accessed: 2016-03-01)<br>http://www.co.washington.mn.us/index.aspx?NID=1606 (last accessed: 2022-03-02)<br>https://city-tampa.opendata.arcgis.com/datasets/building-footprint (last accessed: 2017-12-01)<br>https://data.cityofboston.gov/Permitting/Property-Assessment-2015/yv8c-t43q (last accessed: 2016-03-01)<br>https://gisdata.mn.gov/dataset/us-mn-co-dakota-plan-parcels (last accessed: 2022-03-02)<br>https://gisdata.mn.gov/dataset/us-mn-co-dakota-struc-propertyinfo-buildingp (last accessed: 2022-03-02)<br>https://gisdata.mn.gov/dataset/us-mn-state-metrogis-plan-regonal-prcls-open (last accessed: 2016-03-01)<br>https://gis-monmouthnj.opendata.arcgis.com/datasets/0d2bb2c31e854819939cae0e6e1b589b_0 (last accessed: 2018-12-02)<br>https://gis-monmouthnj.opendata.arcgis.com/datasets/fec0b3d813174cdfb766134315120460_0/ (last accessed: 2022-03-02)<br>https://koordinates.com/layer/102872-sarasota-county-florida-building-footprint/ (last accessed: 2022-03-02)<br>https://mecklenburgcounty.exavault.com/share/view/mg5f-3uke3hyw (last accessed: 2022-03-02)<br>https://opendata-bc-gis.hub.arcgis.com/datasets/building-footprints/ (last accessed: 2016-10-01)<br>https://opendata-bc-gis.hub.arcgis.com/datasets/parcels/explore (last accessed: 2022-03-02)<br>https://public-manateegis.opendata.arcgis.com/datasets/building-footprints (last accessed: 2017-07-01)<br>https://www.bouldercounty.org/government/open-data/ (last accessed: 2022-03-02)<br>https://www.bouldercounty.org/government/open-data/) (last accessed: 2022-03-02)<br>https://www.mass.gov/get-massgis-data (last accessed: 2022-03-02) |
| **Data accessibility** | Repository name: Mendeley Data<br>Data identification number: 10.17632/w33vbvjtdy<br>Direct URL to data: https://data.mendeley.com/datasets/w33vbvjtdy |
| **Related research article** | Leyk, S., Uhl, J. H., Balk, D., & Jones, B. (2018). Assessing the accuracy of multi-temporal built-up land layers across rural-urban trajectories in the United States. Remote Sensing of Environment, 204, 898-917.<br>https://doi.org/10.1016/j.rse.2017.08.035 |